\documentclass[11pt]{article}
\usepackage{amsbsy}
\usepackage{amsmath}
\usepackage{amssymb}
\usepackage{graphicx}
\usepackage{dcolumn}
\usepackage{bm}
\usepackage{psfrag}

\begin{document}
\newcommand{\bbox}[1]{{\bm #1}}
\title{\Large\bf Generalised Ornstein-Uhlenbeck processes}
\author{{\it  V. Bezuglyy$^{\ast }$, B. Mehlig$^{\ast }$,
M. Wilkinson$^{\ast\ast }$, K. Nakamura$^{\dagger}$, \mbox{\rm and}
E. Arvedson$^{\ast }$,
}
\\ [3mm]
\normalsize ($^{\ast }$)
Department of Physics, G\"oteborg University, 41296 Gothenburg, Sweden\\
\normalsize ($^{\ast\ast }$) Faculty of Mathematics and Computing,
The Open University, \\Walton Hall, Milton Keynes, MK7 6AA, England\\
\normalsize ($^{\dagger }$) Department of Applied Physics, Osaka City University,
Osaka 558-8585, Japan}
\vspace{3mm}
\date{}
\maketitle
\par
\vspace{2cm}
 \begin{center}
 {\large\bf Abstract}
 \end{center}
We solve a physically significant extension of a classic problem
in the theory of diffusion, namely the Ornstein-Uhlenbeck process
[G. E. Ornstein and L. S. Uhlenbeck, Phys. Rev. {\bf 36}, 823,
(1930)]. Our generalised Ornstein-Uhlenbeck systems include a force
which depends upon the position of the particle, as well as upon
time. They exhibit anomalous diffusion at short times, and
non-Maxwellian velocity distributions in equilibrium. Two
approaches are used. Some statistics are obtained from a
closed-form expression for the propagator of the Fokker-Planck
equation for the case where the particle is initially at rest. In
the general case we use spectral decomposition of a Fokker-Planck
equation, employing nonlinear creation and annihilation operators
to generate the spectrum which consists of two staggered ladders.
\par
\vspace{1.6cm}
\newpage
\newpage
\section{Introduction}
\begin{figure}[t]
\includegraphics[width=6cm,clip]{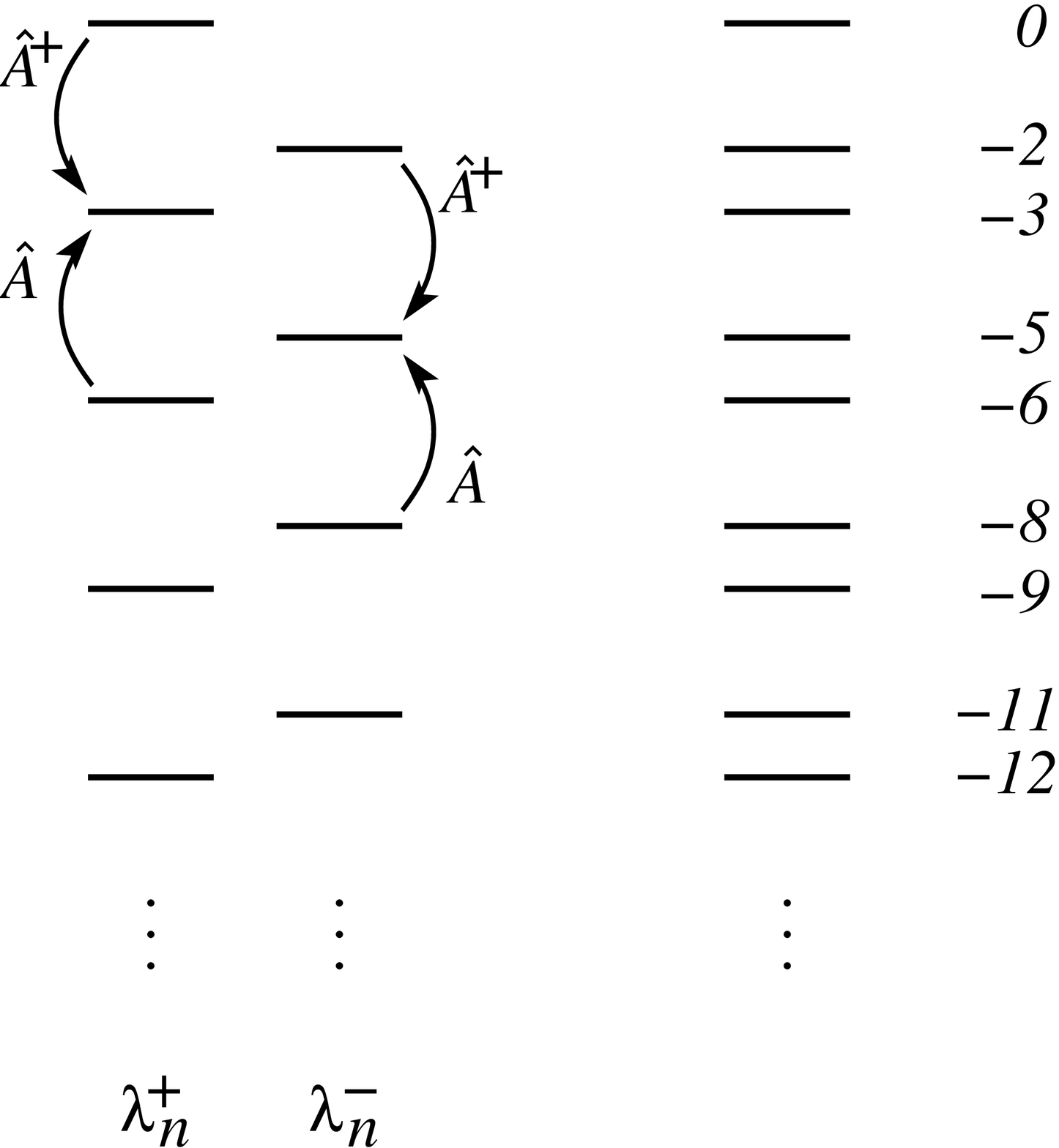}
\caption{\label{fig:1} The spectrum of $\hat H$ consists of two
equally spaced (ladder) spectra $\lambda_n^-$ and $\lambda_n^+$
which are \lq staggered' (that is, they are interleaved with
un-even spacings). $\hat A$ and $\hat A^+$ do not change the
parity of the eigenfunctions.}
\end{figure}

This paper introduces a physically important extension of a
classic problem in the theory of diffusion, namely the
Ornstein-Uhlenbeck process \cite{Orn30}. Our results are obtained
by spectral decomposition of a linear operator. The spectrum of
this operator consists of two ladders of eigenvalues 
with respectively odd and even parity.
The
ladders of eigenvalues are staggered, that is the odd-even step is
different from the even-odd step (see Fig. \ref{fig:1}). 
The corresponding eigenfunctions are generated by a raising operator.
A concise
account of our work on these staggered ladder spectra appeared
earlier \cite{Arv06}. In the following we show how the results
summarised in \cite{Arv06} were obtained.  We also derive new
results, not included in our earlier report: a
closed-form solution for example, and the generalisation 
of our previous results
to a continuous family of diffusion processes.

\subsection{The Ornstein-Uhlenbeck process}
Before we discuss our extension of the Ornstein-Uhlenbeck process,
we describe its usual form \cite{Orn30}. This considers a particle
of momentum $p$ subjected to a rapidly fluctuating random force
$f(t)$ and subject to a drag force $-\gamma p$, so that the
equation of motion is
\begin{equation}
 \label{eq:-1}
 \dot p=-\gamma p+f(t)\,.
\end{equation}
The random force has statistics $\langle f(t)\rangle=0$, $\langle
f(t)f(t')\rangle=C(t-t')$ (angular
 brackets denote ensemble averages throughout).
If the correlation time $\tau$ of $f(t)$ is sufficiently short
($\gamma\tau \ll 1$), the equation of motion may be approximated
by a Langevin equation:
\begin{equation}
\label{eq:langevin} {\rm d}p=-\gamma p{\rm d}t+{\rm d}w\,,
\end{equation}
where the Brownian increment ${\rm d}w$ has statistics $\langle
{\rm d}w\rangle=0$ and $\langle {\rm d}w^2\rangle=2D_0{\rm d}t$.
The diffusion constant is
\begin{equation}
D_0={1\over 2}\int_{-\infty}^\infty{\rm d}t\ \langle
f(t)f(0)\rangle\,.
\end{equation}
This problem is discussed in many textbooks (for example
\cite{vKa81}).

\subsection{Generalised Ornstein-Uhlenbeck processes}
Our extension arises when the force depends upon position as well
as time. We consider the case where the fluctuations of the force
on the particle are mainly a consequence of the spatial, rather
the temporal, fluctuations of the force $f(x,t)$. A consequence of
this difference is that the impulse $\delta w$ supplied to the
particle in a short time $\delta t$ depends upon the momentum of
the particle. If the particle is at position $x_0$ at time $t_0$,
this impulse is
\begin{equation}
\label{eq:dw} \delta w=\int_{t_0}^{t_0+\delta t}\!\!\!{\rm d}t\,
f\big(x_0+p(t-t_0)/m,t\big)+O(\delta t^2)\,.
\end{equation}
In particular, the impulse approaches zero as the speed $\vert
p\vert/m$ of the particle increases, because the motion of the
particle effects an average over the spatial fluctuations of the
force. This can be seen clearly by considering the second moment
of $\delta w$. We assume that the force $f(x,t)$ has the following
statistics
\begin{equation}
\label{eq: stats}
\langle f(x,t)\rangle=0 \,, \ \ \
 \langle f(x,t)f(x',t')\rangle=C(x-x',t-t')\,.
\end{equation}
The spatial and temporal correlation scales of the random force
$f(x,t)$ are $\xi$ and $\tau$ respectively. We consider the case
where (for most of the time) the momentum of the particle is large
compared to $p_0=m\xi/\tau$, then the force experienced by the
particle decorrelates more rapidly than the force experienced by a
stationary particle. If $\delta t$ is large compared to $\tau$ but
small compared to $1/\gamma$, we can estimate the variance of the
impulse $\langle \delta w^2\rangle=2D(p)\delta t$ as follows (due
to translational invariance, we consider without loss of
generality a particle which starts from position $x=0$ at time
$t=0$)
\begin{eqnarray}
\langle \delta w^2\rangle
&=&\int_0^{\delta t}{\rm d}t_1\int_0^{\delta t}{\rm d}t_2\
\big\langle f(pt_1/m,t_1)f(pt_2/m,t_2)\big\rangle \nonumber \\
&=&\delta t\int_{-\infty}^\infty {\rm d}t\ C(pt/m,t)+O(\tau^2)\,.
\end{eqnarray}
We define the momentum diffusion constant by writing $\langle
\delta w^2\rangle=2D(p)\delta t+O(\delta t^2)$, and find
\begin{equation}
\label{eq: 2} D(p)=\frac{1}{2}\int_{-\infty}^\infty {\rm d}t\
C(pt/m,t) \ .
\end{equation}
When $p\ll p_0$ we recover $D(p)=D_0$. When $p\gg p_0$, we can
approximate (\ref{eq: 2}) to obtain
\begin{equation}
\label{eq: 3} D(p)={D_1 p_0\over \vert p\vert}+O(p^{-2}) \ ,\
D_1={m\over 2p_0}\int_{-\infty}^\infty\!\!\!\! {\rm d}X\ C(X,0) \
.
\end{equation}
When the force is the gradient of a potential $V(x,t)$ with a
correlation function having continuous derivatives, we find that
$D_1$ is zero. This case is discussed in section \ref{sec:gen},
where it is shown that $D(p)\sim \vert p\vert^{-3}$ provided the
correlation function of $V(x,t)$ is sufficiently differentiable.
Another variation, also discussed in section \ref{sec:gen}, arises
when the correlation function of the force exhibits a
discontinuity at $t=0$ (as when the potential $V(x,t)$ is itself
generated by an Ornstein-Uhlenbeck process). In this case
$D(p)\sim \vert p\vert^{-2}$, and other exponents are also
possible. We therefore consider a general situation where
$D(p)\sim \vert p\vert^{-\zeta}$ and give exact results for case
\begin{equation}
\label{eq: 4} D(p)=D_\zeta\, (p_0/\vert p|)^\zeta
\end{equation}
with $\zeta\ge 0$. We analyse the dynamics by solving a
Fokker-Planck equation which determines the probability density
for the Langevin process in which the momentum has diffusion
constant given by (\ref{eq: 4}). We discuss the form of this
Fokker-Planck equation in section \ref{sec:FP}; the remainder of
this introduction will set our work in context with earlier
research on related topics.

\subsection{Earlier work}
The motion of a damped particle subjected to a force fluctuating
in both space and time was first studied by Deutsch \cite{Deu85},
who addressed  an entirely different aspect of the problem.
Deutsch considered to case where the momentum of the particle
remains small compared to $p_0$, and posed the question of whether
particles aggregate. He discovered that there is a phase
transition between coalescing and non-coalescing trajectories.
(Two of the authors of the present paper subsequently solved
Deutsch's one-dimensional model exactly \cite{Wil03}, and  results for
two and three spatial dimensions are discussed in
\cite{Meh04,Dun05}). All of these papers only considered cases
where $p\ll p_0$.

Sturrock \cite{Stu66} analysed the motion of a  particle subjected
to a spatially varying force field without damping. He  introduced
the concept of a momentum diffusion constant which varies as a
function of the momentum: that is, he considered the same problem
as is addressed in the present paper, but in the limit of damping
constant $\gamma=0 $. Subsequently Golubovic, Feng, and Zeng
\cite{Gol91} identified the importance of the relation $D(p)\sim
\vert p\vert^{-3}$ (in the case of a potential force), and
discussed the nature of the Fokker-Planck equation and its
solution in the case where $\gamma =0$. It was argued that the
particle exhibits anomalous diffusion and solution for the
propagator of the Fokker-Planck equation with initial value $p=0$
was proposed. Later Rosenbluth \cite{Ros92} pointed to an error in
the evaluation of this propagator. The  results
of \cite{Stu66,Gol91,Ros92} were applied to
the stochastic acceleration \cite{Fer49} of particles in plasmas,
and subsequent contributions have concentrated on refining models
for the calculation of $D(p)$ (see, for example,
\cite{Ach91,Sta04}).

In the following we analyse the problem with the damping term,
proportional to $\gamma$, included. Surprisingly, we find that
this more general problem is more tractable: we are able, for
example, to obtain precise results concerning the problems
considered in \cite{Gol91,Ros92} by taking, in our solutions, 
the limit $\gamma \to 0$.

There is a large literature devoted to the motion of particles
advected in random velocity fields (corresponding to the
large-$\gamma$ limit of the model we study). In the case where the
velocity field is independent of time, sub-diffusive motion is
typically found \cite{Sin80,Isi92}. The advection of tracers in a
turbulent fluid is described by models with rapidly fluctuating
velocity fields \cite{Fal01}. In our problem the inertia of the
particles plays an important role. Particles suspended in a
turbulent fluid can show surprising clustering properties when
inertia effects are significant. These were first proposed by
Maxey \cite{Max87}; the current state of knowledge is summarised
in \cite{Dun05}. In cases where the random force results from
motion of the surrounding fluid, it is not possible for the
condition $p\gg p_0$ to be realised \cite{Dun05}.

We remark that a brief summary of many of the results of this
paper was already published \cite{Arv06}. The closed-form solution
of section \ref{sec:CF}, the WKB analysis and most of the results
for general values of $\zeta $ were not discussed in this earlier
work.

\subsection{Description of our results and outline of this paper}

In order to simplify the presentation, we describe in detail only
our results for the case $\zeta=1$, corresponding to a generic
random force. Corresponding expressions for general values of $\zeta$ are
obtained using the same method, and we quote the most important
results for general values of $\zeta$ in section \ref{sec:gen} at the end of
the paper.

In section \ref{sec:FP}, the Fokker-Planck equation for the
generalised Ornstein-Uhlenbeck processes is described. In section
\ref{sec:CF} we briefly discuss a particular closed-form solution,
which enables us to determine the steady state momentum
distribution (which is non-Maxwellian) and some statistics, such
as the time evolution of the variance of the momentum. The results
of section \ref{sec:CF} are not sufficient to enable all
statistics to be calculated, and in the general case we obtain
statistics via a spectral decomposition of the Fokker-Planck
equation. Section \ref{sec:specdec} discusses this spectral
decomposition. We transform the Fokker-Planck operator into a
Hermitean operator and determine the eigenvalues and eigenvectors
of this \lq Hamiltonian' operator by generating them using a new
type of raising and lowering operators, which are nonlinear
second-order differential operators. We show that the resulting
spectrum is a ladder spectrum, consisting of separate ladders for
the odd and even parity states. These are staggered: the odd-even
separation differs from even-odd. Section \ref{sec:corr} contains
calculations of the matrix elements needed for computing
correlation functions and expectation values. In section
\ref{sec:diff} we summarise our results on diffusion and anomalous
diffusion for generic random forcing.

Section \ref{sec:wkb} discusses a technical issue concerning our
evaluation of the spectrum. When the index of the eigenvalue is
large, it is possible to apply standard WKB approximation methods
everywhere except in the vicinity of a singularity of the
Hamiltonian. We show that the singularity introduces phase shifts
which explain the staggered-ladder structure of the spectrum.

Finally, in section \ref{sec:gen} we explain in more detail how
other values of $\zeta$ can arise and summarise our results for
general $\zeta $.

\section{Fokker-Planck equations}
\label{sec:FP}

We consider a particle with equations of motion
\begin{equation}
\dot x=p/m\ ,\ \ \ \dot p=-\gamma p+f(x,t)
\end{equation}
where the force $f(x,t)$ is random, with statistics given by
equation (\ref{eq: stats}). In the limit as the correlation time
$\tau$ of the force approaches zero, the equation of motion of the
momentum may be approximated by a Langevin equation,
(\ref{eq:langevin}), where the random increment ${\rm d}w$ has
second moment $\langle {\rm d}w^2\rangle=2D(p){\rm d}t$ with
$D(p)$ given by (\ref{eq: 2}). This Langevin equation for the
stochastic evolution of $p(t)$ corresponds to a Fokker-Planck
equation (generalised diffusion equation) for the probability
density of the momentum, $P(p,t)$. Using standard results
\cite{vKa81}, the Fokker-Planck equation is
\begin{equation}
{\partial P\over{\partial t}}=-{\partial\over{\partial
p}}\bigl(v(p)P\bigr)+{\partial^2\over{\partial^2
p}}\bigl(D(p)P\bigr)
\end{equation}
where
\begin{equation}
v(p)={\langle {\rm d}p\rangle\over{{\rm d}t}}\ ,\ \ \
D(p)={\langle {\rm d}p^2\rangle\over{2{\rm d}t}} \ .
\end{equation}
Note that we can replace ${\rm d}p$ by ${\rm d}w$ in the expression
for $D(p)$, because the neglected terms are of higher order in
${\rm d}t$, and that $D(p)$ has already been obtained in equation
(\ref{eq: 2}). In order to determine the correct form of the
Fokker-Planck equation it remains to determine $\langle {\rm
d}p\rangle=-\gamma {\rm d}t +\langle {\rm d}w\rangle$.

Expanding the impulse (\ref{eq:dw}) about a reference trajectory
$x(t)=pt/m$, and using the fact that $\langle f(x,t)\rangle=0$, we
obtain
\begin{eqnarray}
\langle \delta w\rangle&=&{1\over m} \int_0^{\delta t}\!\!\!{\rm d}t_1
\int_0^{t_1}\!\!\!{\rm d}t_2\int_0^{t_2}\!\!\!{\rm d}t_3\
\exp[-\gamma(t_2-t_3)]
\biggl\langle {\partial f\over{\partial x}}(pt_1/m,t_1)f(pt_3/m,t_3)
\biggr\rangle\,.
\label{eq: 2.3b}
\end{eqnarray}
Note that throughout the three-dimensional region of integration,
we have $0\le t_3 \le t_2 \le t_1 \le \delta t$, and the short
correlation time implies that the integrand is negligible unless
$\vert t_1-t_3\vert < \tau$. The integrand is therefore
significant along a line rather than a surface, because $t_2$ must
lie between $t_1$ and $t_3$. The integral is therefore $O(\delta
t)$, rather than $O(\delta t^2)$ which would obtain if the
integrand were significant on a surface. We replace the factor
$\exp[-\gamma (t_2-t_3)]$ by unity because $\gamma \tau \ll 1$,
and the other factor is negligible when $\vert t_2-t_3\vert >
\tau$. The integral over $t_2$ then gives simply $t_1-t_3$.
Writing $t=t_1-t_3$, in the limit $\gamma \tau \ll 1$, the result
is therefore
\begin{eqnarray}
\langle \delta w\rangle&=&{1\over {2m}}\int_{-\infty}^\infty {\rm d}t\
t\,\biggl\langle {\partial f\over{\partial
x}}(0,0)f(pt/m,t)\biggr\rangle= \delta t\,\frac{{\rm d}}{{\rm d}p}D(p)\,.  \label{eq: av}
\end{eqnarray}
This implies
\begin{equation}
\label{eq:dp}
v(p) = \frac{\rm d}{{\rm d}p} D(p)\,.
\end{equation}
Rosenbluth \cite{Ros92} has pointed out that this relation can
also be obtained as a consequence of applying the principle of
detailed balance.

With (\ref{eq: 2}) and (\ref{eq:dp}), the following Fokker-Planck
equation obtains:
\begin{equation}
\label{eq:FP}
{\partial P\over{\partial t}}={\partial\over{\partial p}}\big(\gamma p
+D(p){\partial\over{\partial p}}\big)P\,.
\end{equation}
Sturrock \cite{Stu66} introduced a related Fokker-Planck equation
(without the damping term) and also gave an expression for $D(p)$
analogous to equation (\ref{eq: 2}).

In the following we discuss our solution of (\ref{eq:FP}) for the particular
case of generic random forcing (corresponding to $\zeta = 1$). Results for
other values of $\zeta$ are obtained in an analogous fashion. The
general case is briefly described in section \ref{sec:gen}.

\section{A particular closed-form solution}
\label{sec:CF}
In this section we introduce a particular solution of the Fokker-Planck
equation (\ref{eq:FP}) with $D(p)$ given by (\ref{eq: 4}). 
We restrict ourselves to
the case of generic random forcing
(corresponding to $\zeta=1$):
\begin{equation}
\label{eq:FP1}
{\partial P\over{\partial t}}={\partial\over{\partial p}}\big(\gamma p
+D_1 \frac{p_0}{|p|}{\partial\over{\partial p}}\big)P\,.
\end{equation}
Consider the distribution $P(p,t)$ of momentum $p$ for particles
initially at rest. It satisfies the initial condition $P(p,0) =
\delta(p)$ where $\delta(p)$ is the $\delta$-function. For this
particular initial condition, we have found the following
closed-form solution of (\ref{eq:FP1}):
\begin{equation}
\label{eq:partsol}
P(p,t)=\frac{1}{2\Gamma(4/3)}\,\frac{\gamma^{1/3}}{[3p_0D_1(1-{\rm
e}^{-3\gamma t})]^{1/3}} {\rm exp}\left [-\frac{\gamma
|p|^3}{3p_0D_1(1-{\rm e}^{-3\gamma t})} \right ]\,.
\end{equation}
Eq. (\ref{eq:partsol}) determines how the moments of momentum grow
for a particle initially at rest:
\begin{equation}
\label{eq:pm2l}
\langle p^{2l}(t)\rangle = \Big(\frac{3 D_1}{\gamma}\Big)^{2l/3}
\frac{\Gamma\big((2l+1)/3\big)}{\Gamma(1/3)}
\big(1-{\rm e}^{-3\gamma t}\big)^{2l/3}
\end{equation}
for positive integers $l$. This result is consistent with the
result obtained in \cite{Arv06} (eq. (8) in that paper). In the
limit of small times (\ref{eq:pm2l}) gives rise to anomalous
diffusion
\begin{equation}
\label{eq:p2l}
\langle p^{2l}(t)\rangle \sim t^{2l/3}\,.
\end{equation}
At large times ($\gamma t\gg 1$), by contrast,  we  obtain a stationary
non-Maxwellian momentum distribution
\begin{equation}
\label{eq: 9} P_0(p) =
\frac{1}{2\Gamma(4/3)}\,\frac{\gamma^{1/3}}{(3p_0D_1)^{1/3}}
\exp\big[-\gamma |p|^3/(3 p_0 D_1)\big]\,.
\end{equation}
The particular solution (\ref{eq:partsol}) generalises in a natural
way to other values of $\zeta$.

However, in order to determine the momentum correlation function
and the spatial diffusion properties, the particular solution
(\ref{eq:partsol}) is not sufficient, the general solution for
arbitrary initial condition is required. We have not been able to
obtain the general solution to (\ref{eq:FP}) in closed form.
Therefore, we determine it using spectral decomposition: we
construct the eigenvalues $\lambda_n$ and eigenfunctions
$\psi_n(z)$ of a Hermitian operator $\hat H$ corresponding to the
Fokker-Planck equation (\ref{eq:FP1}). We identify raising and
lowering operators $\hat A^+$ and $\hat A$ which map one
eigenfunction to another with respectively two more or two fewer
nodes. We use these to obtain the spectrum and eigenfunctions of
$\hat H$ which in turn allow us to construct the propagator,
expectation values and correlation functions. This approach is
described in sections \ref{sec:specdec} to \ref{sec:corr}.

\section{Spectral decomposition}
\label{sec:specdec}
Introducing dimensionless variables ($t' = \gamma t$ and
$p =  z p_0 [D_1/(\gamma p_0^2)]^{1/3}$) we write (\ref{eq:FP1}) as
\begin{equation}
\label{eq:3}
{\partial P\over{\partial t'}}={\partial\over{\partial z}}\big(z
+\frac{1}{|z|}{\partial\over{\partial z}}\big)P\equiv \hat F P
\end{equation}
It is convenient to transform the Fokker-Planck operator
$\hat F$
to a Hermitian form which we shall refer to as the Hamiltonian operator:
     \begin{equation}
       \label{eq:H}
         \hat H = P_0^{-1/2}\hat F P_0^{1/2} =
    \frac{1}{2}-\frac{|z|^3}{4}+\frac{\partial}{\partial z}
     \frac{1}{|z|}
 \frac{\partial}{\partial z}\,.
 \end{equation}
Here $P_0(z)  \propto \exp(-|z|^3/3)$ is the stationary solution (\ref{eq: 9})
satisfying $\hat F P_0= 0$.  We solve the diffusion problem by constructing the
eigenfunctions of the Hamiltonian operator. In the following we make
use of Dirac notation \cite{Dir30} of quantum mechanics to write the equations in a
compact form and to emphasise their structure.

The eigenfunctions of the Fokker-Planck equation (\ref{eq:FP}) are
alternately even and odd functions, defined on the interval
$(-\infty,\infty)$. The operator $\hat H$, describing the limiting
case of this Fokker-Planck operator, is singular at $z=0$. We
identify two eigenfunctions of $\hat H$ by inspection,
$\psi_0^+(z) = {\cal C}_0^+\,\exp(-|z|^3/6)$ which has eigenvalue
$\lambda_0^+ = 0$ and $\psi_0^-(z) = {\cal C}_0^-\, z |z|
\exp(-|z|^3/6)$ with eigenvalue $\lambda^-_0 = -2$. These
eigenfunctions are of even and odd parity, respectively (zero and
one node, respectively). Our approach to determining the full
spectrum is to define a raising operator $\hat A^+$ which maps any
eigenfunction $\psi_n^\pm(z)$ to its successor with the same
parity, $\psi_{n+1}^\pm(z)$, having two additional nodes.

\subsection{Algebra of raising and lowering operators}
We write
\begin{equation}
\label{eq:ham}
\hat H = \hat a^- |z|^{-1} \hat a^+\,.
\end{equation}
Here $a^\pm = (\partial_z \pm z |z|/2)$. We introduce the
operators
\begin{equation}
\label{eq:rl}
\hat A = \hat a^+ |z|^{-1} \hat a^+\quad\mbox{and}\quad
\hat A^+ = \hat a^- |z|^{-1} \hat a^-
\end{equation}
as well as
\begin{equation}
\label{eq:gam}
\hat G = \hat a^+ |z|^{-1} \hat a^-\,.
\end{equation}
Note that $\hat A^+$ is the Hermitian conjugate of $\hat A$.
The commutator of $\hat A$ and $\hat A^+$ is
\begin{equation}
[\hat A, \hat A^+] = -3(\hat H +\hat G)\,.
\end{equation}
Note also that $\hat H - \hat G = \hat I$ (where $\hat I$ is the
identity operator).

\subsubsection{Eigenvalues}
It can be verified that
\begin{equation}
[\hat H, \hat A] = 3\hat A\quad \mbox{and}\quad
[\hat H, \hat A^+] = -3\hat A^+\,.
\end{equation}
 These expressions show that
 the action of $\hat A$ and $\hat A^+$ on any eigenfunction is to
 produce another eigenfunction with eigenvalue increased or
 decreased by three, or else to produce a function which is
 identically zero. The operator $\hat A^+$ adds two nodes,
 and repeated action of $\hat A^+$ on $\psi_0^+(z)$ and
 $\psi_0^-(z)$ therefore exhausts the
 set of eigenfunctions. Together with $\lambda_0^+ = 0$ and
 $\lambda_0^- = -2$ this establishes that the spectrum of $\hat H$
 is (see Fig. \ref{fig:1})
 \begin{eqnarray}
 \label{eq:32}
 \lambda_n^+ &=& -3n\quad\mbox{and}\quad\lambda_n^- = -3n-2\,\quad n =
 0,\ldots,\infty\,.
 \end{eqnarray}

\subsubsection{Eigenfunctions}
\label{sec:eigenstates}
We represent the eigenfunctions by of $\hat H$ by kets
$|\psi_n^-\rangle$ and $|\psi_n^+\rangle$. The actions of $\hat A$
and $\hat A^+$ are
\begin{eqnarray}
\label{eq:C} \hat A^+ |\psi_{n}^\pm\rangle &=& C_{n+1}^\pm
|\psi_{n+1}^\pm \rangle \quad\mbox{and}\quad \hat A
|\psi_{n}^\pm\rangle=C_{n+1}^\pm|\psi_{n-1}^\pm \rangle \,.
\end{eqnarray}
The normalisation factor $C_{n+1}^-$ is determined as follows:
\begin{equation}
1=\langle\psi_{n+1}^-|\psi_{n+1}^- \rangle=
 (C_{n+1}^-)^{-2}  \langle \psi_{n}^-|\hat A\hat A^{+}|\psi_n^-\rangle
=  (C_{n+1}^-)^{-2}\langle \psi_{n}^- | [\hat A, \hat A^+] + \hat A^+\hat
A|\psi_n^-\rangle\,.
\end{equation}
It follows
\begin{equation}
(C_{n+1}^-)^2 = \Big(3\,(-2\lambda_n^-+1)+(C_{n}^-)^2\Big)\,.
\end{equation}
By recursion we obtain
\begin{equation}
\label{eq:Cm}
C_{n}^- = [3n(3n+2)]^{1/2}\,.
\end{equation}
This determines the normalisation of the states
$ (\hat A^{+})^{n} |\psi_{0}^- \rangle $
\begin{equation}
\label{eq:A1}
|\psi_{n}^- \rangle =
N_{n}^- (\hat A^{+})^{n} |\psi_{0}^-
\rangle
\end{equation}
with
\begin{equation}
N_{n}^-=\left(\prod_{k=1}^n 3k(3k+2)\right)^{-1/2}\,N_0^-\,.
\end{equation}
For the positive-parity states we proceed in a similar fashion and obtain
\begin{equation}
\label{eq:Cp}
C_{n}^+ = [3n(3n-2)]^{1/2}\,.
\end{equation}
This implies
\begin{equation}
\label{eq:A3}
|\psi_{n}^+ \rangle = N_{n}^+ (\hat A^{+})^{n} |\psi_{0}^+
\rangle
\end{equation}
with
\begin{equation}
N_{n}^+ =
\left(\prod_{k=1}^n 3k(3k-2)\right)^{-1/2} \,N_0^+\,.
\end{equation}
The operators $\hat A^+$ and $\hat A$ differ from the usual
examples of raising and lowering operators in that they are of
second order in ${\rm d}/{\rm d} z$, whereas other examples of
raising and lowering operators are of first order in the
derivative. The difference is associated with the fact that the
spectrum is a staggered ladder: only states of the same parity
have equal spacing, so that the raising and lowering operators
must preserve the odd-even parity. This suggests replacing a
first-order operator which increases the quantum number (total
number of nodes) by one with a second-order operator which
increases the quantum number by two, preserving parity.

There is an alternative approach to generating the eigenfunctions
of $\hat H$. This equation falls into one of the classes
considered in \cite{Inf51}, and we have written down first-order
operators which map one eigenfunction into another. However, these
operators are themselves functions of the quantum number $n$,
making the algebra cumbersome. The approach is briefly described
in the next section.

\subsection{Schr\"odinger factorisation}
\label{sec:fac}
Consider the eigenvalue problem
\begin{equation}
\hat{H} |\psi\rangle = \lambda |\psi\rangle
\label{eigenv}
\end{equation}
with Hamiltonian (\ref{eq:H}). For $z > 0$ it can be transformed
by the variable change $x=z^3$:
\begin{equation}
\label{eq:F}
\left[(3x)^2\frac {{\rm d}^2}{{\rm d}x^2}+3x\frac {{\rm d}}{{\rm d}x}
-x\bigg(\frac {1}{4}x+\lambda-\frac {1}{2}\bigg)\right]\psi(x)= 0
\end{equation}
by means of the transformation $x=z^3$.
Eq. (\ref{eq:F}) is a Fuchsian
linear differential equation  with
regular singular points of rank less than or equal to two.
Eq. (\ref{eq:F}) can therefore be factorised using
a generalised Schr\"odinger factorisation scheme (see \cite{Inf51} for
a review of this method).

Applying this scheme we have obtained raising and lowering operators
generating the spectrum (\ref{eq:32}).
The raising operator acting on $\psi_n^\pm$ is given by
\begin{equation}
T_{\pm, n+1}=-3x\frac{{\rm d}}{{\rm d}x}+\frac{x}{2}-3(n+\delta _\pm )
\end{equation}
with $\delta_+=1/3$ and $\delta _-=1$.
The lowering operator acting on $\psi_n^\pm$ is
\begin{equation}
\tilde T_{\pm, n}=-3x\frac{{\rm d}}{{\rm d}x}-\frac{1}{2}x+3(n+\eta _\pm)
\end{equation}
with $\eta_+=0$ and $\eta _{-}=2/3$.
Note that the Hermitian conjugates of $T_{\pm, n+1}$  and
$\tilde T_{\pm, n}$ are
\begin{eqnarray}
(T_{\pm, n+1})^+ &=&-\tilde T_{\pm, n+1}+3\\
(\tilde T_{\pm, n})^+ &=&-T_{\pm, n}+3\,.
\end{eqnarray}
The raising and lowering operators satisfy
\begin{eqnarray}
T_{\pm,n+1} |\psi_n^\pm\rangle = C_{n+1}^\pm |\psi_{n+1}^\pm\rangle\quad
\mbox{and}\quad \tilde T_{\pm,n} |\psi_n^\pm\rangle = C_{n}^\pm
|\psi_{n-1}^\pm\rangle
\end{eqnarray}
with $C_n^\pm$ given by (\ref{eq:Cm}) and (\ref{eq:Cp}),
generating the spectrum (\ref{eq:32}). The operators differ from
$\hat A$ and $\hat A^+$ introduced in section \ref{sec:specdec} in
that they are of first order in ${\rm d}/{\rm d}x$, and in that
they depend on the state they are applied to.

\section{Correlation functions and matrix elements}
\label{sec:corr}
\subsection{Correlation functions}

The required solutions of the Fokker-Planck equation may be
expressed in terms of the propagator $K(y,z,t)$ which is the
probability density for the scaled momentum to reach $z$ after
time $t$, starting from $y$. It satisfies the Fokker-Planck
equation $\partial_{t'} K = \hat F K$ and can be expressed in
terms of the eigenvalues $\lambda_n^\sigma$ and eigenfunctions
$\phi_n^\sigma(z) = P_0^{-1/2} \psi_n^\sigma(z)$ of $\hat F$:
\begin{equation}
K(y,z;t') = \sum_{n\sigma} a_n^\sigma(y) \phi_n^\sigma(z)
\exp(\lambda_n^\sigma t')
\end{equation}
for $t'>0$.
Here $y$ is the initial value and $z$ is the final value of the
coordinate. The expansion coefficients $a_n^\sigma(y)$ are
determined by the initial condition $K(y,z;0) = \delta(z-y)$,
namely $a_n^\sigma(y) = P_0^{-1/2}\psi_n^\sigma(y)$. In terms of
the eigenfunctions of $\hat H$ we have (for $t'>0$)
\begin{equation}
K(y,z;t') = \sum_{n\sigma} P_0^{-1/2}(y) \psi_n^\sigma(y)
      P_0^{1/2}(z)\psi_n^\sigma(z)
\exp(\lambda_n^\sigma t')\,.
\end{equation}
Equilibrium correlation functions of an observable
$O(z)$ are given by
\begin{equation}
\label{eq:eqc} \langle O(z_0) O(z_{t'})\rangle_{\rm eq.} =
\int_{-\infty}^\infty\!{\rm d}z\int_{-\infty}^\infty\! {\rm d}y\
O(z) O(y) K(y,z;t') P_0(y)\,.
\end{equation}
Since $P_0^{1/2}(y) = \psi_0^+(y)$ this corresponds to  
\begin{equation}
\label{eq:OO}
\langle O(z_0) O(z_{t'})\rangle_{\rm eq.}
= \sum_{n\sigma} |\langle \psi_0^+|\hat O|\psi_n^\sigma\rangle|^2
\exp(\lambda_n^\sigma t')
\end{equation}
 for $t'>0$.
The momentum correlation function in equilibrium, for instance, is
\begin{equation}
\label{eq:pcorr}
\langle p_t p_0\rangle_{\rm eq.} =
 p_0^2\left(\frac{D_1}{\gamma p_0^2}\right)^{2/3}
\sum_{n} \langle \psi_0^+|\hat z|\psi_n^-\rangle^2 \exp(\lambda_n^- t')
\end{equation}
 for $t'>0$,
which requires the evaluation of matrix elements $\langle
\psi_0^+\vert\hat z\vert\psi_n^-\rangle$.

Consider on the other hand the time-dependence of $\langle
x^2(t)\rangle$, with particles initially at rest at the origin. We
need to evaluate
\begin{equation}
\langle x^2(t)\rangle = \frac{1}{m^2}\int_0^t\!{\rm d}t_1
\int_0^t\!{\rm d}t_2\  \langle p_{t_1} p_{t_2}\rangle\,.
\end{equation}
In dimensionless variables this corresponds to
\begin{eqnarray}
\label{eq:60} \langle x^2(t)\rangle &=& \frac{1}{\gamma^2}
\left(\frac{p_0 D_1}{\gamma}\right)^{2/3} \frac{1}{m^2}
\int_0^{t'}\!{\rm d}t_1'\int_0^{t'}\!{\rm d}t_2'\ \langle z_{t_1'}
z_{t_2'}\rangle\,.
\end{eqnarray}
The required correlation function is (assuming $t_2'> t_1'>0$)
\begin{eqnarray}
\langle z_{t_2'} z_{t_1'}\rangle &=& \int_{-\infty}^\infty\!{\rm
d}z_1\int_{-\infty}^\infty\!{\rm d}z_2\ z_1 z_2\,
K(z_1,z_2;t_2'-t_1')\,
K(0,z_1;t_1')\nonumber\\
&=&  \sum_{n,m} \frac{\psi_m^+(0)}{ \psi_0^+(0)}
\langle \psi_{0}^+ |\hat z|\psi_{n}^{-}\rangle
\langle \psi_{n}^- |\hat z|\psi_{m}^{+}\rangle
\exp[\lambda_{n}^-(t_2'-t_1')+\lambda_m^+ t_1']\,.
\label{eq:61}
\end{eqnarray}
In order to evaluate (\ref{eq:61}), the ratios of wave-function
amplitudes ${\psi_m^+(0)}/{ \psi_0^+(0)}$ are required in addition
to matrix elements of $\hat z$. The matrix elements $\langle
\psi_{n}^- |\hat z|\psi_{m}^{+}\rangle$ are determined in section
\ref{sec:x2}, while the ratios of eigenfunctions are calculated in
section \ref{sec:w}.

\subsection{Matrix elements $\langle \psi_m^+ | \hat z |\psi_n^-\rangle$}
\label{sec:x2}

To evaluate the matrix elements $Z_{mn} =  \langle \psi_{m}^+
|\hat z|\psi_{n}^{-}\rangle$ we proceed in three steps: we first
evaluate $Z_{0n}$, in a second step, the matrix elements $Z_{mn}$
are related to $Z_{0,n-m}$ for $m \leq n$. Third, $Z_{mn}$  is
evaluated for $m>n$.

\subsubsection{Matrix elements $\langle \psi_0^+ | \hat z |\psi_n^-\rangle$}
Consider first $Z_{0n} = \langle \psi_0^+|\hat z|\psi_n^-\rangle$.
These matrix elements are obtained by recursion. To evaluate
\begin{eqnarray}
Z_{0,n+1} &=& \langle \psi_0^+|\hat z \hat
A^+|\psi_n^-\rangle/C_{n+1}^-
\end{eqnarray}
we write $\hat z\hat A^+ = \hat z \hat G + \hat z (\hat A^+-\hat G)
= \hat z (\hat H-\hat I) +\hat z (\hat A^+-\hat G)$. It follows
\begin{equation}
\langle \psi_0^+|\hat z \hat A^+|\psi_n^-\rangle
 = (\lambda_n^--1) Z_{0n} + \langle \psi_0^+| \hat z (\hat A^+-\hat
  G)|\psi_n^-\rangle\,.
  \end{equation}
  Using $(\hat A^+-\hat G) = -\hat z\hat a^- $
  and $[\hat z^2,\hat a^-] = -2\hat z$
  we obtain
  \begin{equation}
  \langle \psi_0^+|\hat z \hat A^+|\psi_n^-\rangle =
  (\lambda_n^-+1) Z_{0n}\,.
  \end{equation}
  This corresponds to the recursion
  \begin{equation}
  Z_{0n} =  (-1)^n
  \frac{\prod_{k=0}^{n-1} (3k+1)}{\sqrt{\prod_{k=0}^{n-1} 3(k+1)(3k+5)}}
  Z_{00}
  \end{equation}
With $Z_{00} = 3^{-5/12}\sqrt{\pi}/\Gamma(2/3)$ (found by direct
evaluation of an  integral) we obtain
  \begin{equation}
  \label{eq:J0n}
  Z_{0n} = (-1)^n \frac{ 3^{-5/12}}{\sqrt{2\pi}} \sqrt{\Gamma(2/3)}
  \frac{\Gamma(n+1/3)}{\sqrt{\Gamma(n+1)\Gamma(n+5/3)}}\,.
  \end{equation}

\subsubsection{Matrix elements $\langle \psi_m^+ | \hat z |\psi_n^-\rangle$
for $m\leq n$}
Consider
now   the case $m \leq n$. Let
\begin{eqnarray}
\nonumber
J_{mn} =
\langle \psi_{0}^+ |\hat A^{m} \hat z (\hat A^+)^{n}|\psi_{0}^{-}\rangle
 =   \langle \psi_{0}^+ |\hat A^{m} [\hat z, \hat A^+]
         (\hat A^+)^{n-1}|\psi_{0}^{-}\rangle
         + \langle \psi_{0}^+ |\hat A^{m}  \hat A^+ \hat z
         (\hat A^+)^{n-1}|\psi_{0}^{-}\rangle\nonumber\,.
\end{eqnarray}
We use
$[\hat z,\hat A^+] = -(|\hat z|^{-1}\hat a^- + \hat a^- |\hat z|^{-1})$
to write
\begin{eqnarray}
J_{mn} &=&   -\langle \psi_{0}^+ |\hat A^{m} \hat
|z|^{-1}\hat a^- (\hat A^+)^{n-1}|\psi_{0}^{-}\rangle
-\langle \psi_{0}^+ |\hat A^{m} \hat a^- |\hat z|^{-1}
                  (\hat A^+)^{n-1}|\psi_{0}^{-}\rangle \nonumber\\
                  &&+ \langle \psi_{0}^+ |\hat A^{m}  \hat A^+ \hat z
         (\hat A^+)^{n-1}|\psi_{0}^{-}\rangle\nonumber\\
&=&   J_{mn}^{(1)}+ J_{mn}^{(2)}+ J_{mn}^{(3)}
\end{eqnarray}
and evaluate the three terms separately. The third one gives
\begin{equation}
J_{mn}^{(3)}=  \langle \psi_{0}^+ |\hat A^{m}  \hat A^+ \hat z
(\hat A^+)^{n-1}|\psi_{0}^{-}\rangle
= (C_m^+)^2 \langle \psi_{0}^+ |\hat A^{m-1}
\hat z (\hat A^+)^{n-1}|\psi_{0}^{-}\rangle = (C_m^+)^2 J_{m-1n-1}\,.
\end{equation}
Consider next the first term:
$\hat A^{m} |\hat z|^{-1}\hat a^- (\hat A^+)^{n-1}
= \hat A^{m-1} \hat a^+ |\hat z|^{-1} \hat a^+ |\hat z|^{-1} \hat a^-
(\hat A^+)^{n-1}
= \hat A^{m-1} \hat a^+ |\hat z|^{-1} \hat G (\hat A^+)^{n-1}$, and
thus
\begin{eqnarray}
\langle \psi_{0}^+ |\hat A^{m} |\hat z|^{-1}\hat a^-
                  (\hat A^+)^{n-1}|\psi_{0}^{-}\rangle
&=& (\lambda_{n-1}^--1)\langle \psi_{0}^+ |\hat A^{m-1} \hat a^+
|\hat z|^{-1} (\hat A^+)^{n-1}|\psi_{0}^{-}\rangle\,.
\end{eqnarray}
Using
$\hat A^{m-1} \hat a^+ |\hat z|^{-1}  (\hat A^+)^{n-1}
= \hat A^{m-1} \hat a^+ |\hat z|^{-1} \hat a^- |\hat z|^{-1}\hat a^-(\hat A^+)^{n-2}
= \hat A^{m-1}\hat G |\hat z|^{-1}\hat a^-\hat
A^{+n-2}$
we obtain the recursion
\begin{eqnarray}
J_{mn}^{(1)} &=& (\lambda_{n-1}^--1) (\lambda_{m-1}^+-1)
J_{m-1n-1}^{(1)}=3n(3m-2)J_{m-1n-1}^{(1)}
\end{eqnarray}
Now consider  $J_{mn}^{(2)}$.  Using
$\hat a^- |\hat z|^{-1} (\hat A^+)^{n-1} = \hat A^{+} |\hat z|^{-1} \hat
a^-\hat A^{+n-2}$ it follows
\begin{eqnarray}
J_{mn}^{(2)} &=& (C_m^+)^2 J_{m-1n-1}^{(1)}\,.
\end{eqnarray}
This implies
\begin{eqnarray}
J_{mn}^{(2)} &=& \frac{(C_m^+)^2}{ (\lambda_{n-1}^--1)
(\lambda_{m-1}^+-1)} J_{mn}^{(1)}
= \frac{m}{n} J_{mn}^{(1)}
\label{eq:63}
\end{eqnarray}
Note also that $J_{0n}^{(3)} = 0$, as well as $J_{0n}^{(2)}= 0$.
This gives  $J_{0n}^{(1)} = J_{0n}$
[consistent with (\ref{eq:63})]. We obtain
\begin{equation}
J_{mn}^{(1)} =\prod_{k=1}^m 3(n-m+k)(3k-2)J_{0n-m}
\end{equation}
which results in
\begin{eqnarray}
\label{eq:65}
J_{mn}^{(1)} &=&
(-1)^{m+n} \frac{\Gamma(2/3)3^{m+n+5/6}}{6\pi}
\frac{\Gamma(n+1)\Gamma(m+1/3)\Gamma(n-m+1/3)}{\Gamma(n-m+1)}\,.
\end{eqnarray}
Eq. (\ref{eq:65})
allows us to write down an inhomogeneous recursion for $J_{mn}$:
\begin{eqnarray}
J_{mn} &=& 3m(3m-2) J_{m-1n-1}+(1+m/n)J_{mn}^{(1)}
\end{eqnarray}
where the inhomogeneous term is given by (\ref{eq:65}).
Iterating this recursion we  obtain
\begin{eqnarray}
J_{mn}
       &=& \sum_{l=0}^m \left(\prod_{k=l+1}^m 3k(3k\!-\!2)\right)
       \big(1\!+\!\frac{l}{n\!-\!m\!+\!l}\big) J_{ln-m+l}^{(1)}
       \end{eqnarray}
       which results in
\begin{equation}
\label{eq:Jmn}
J_{mn} \!=\! (-1)^{m-n} \frac{3^{m+n+1}\Gamma(2/3)^2}{4\pi^2}
(m\!+\!n\!+\!1)
\frac{\Gamma(n+1)\Gamma(m+1/3)\Gamma(n-m+1/3)}{\Gamma(n-m+2)}
J_{00}\,.
\end{equation}
This determines $J_{mn}$ for $n\geq m$.

\subsubsection{Matrix elements $\langle \psi_m^+ | \hat z
|\psi_n^-\rangle$ for $m> n$}
For $m>n$ we use instead
\begin{eqnarray}
J_{mn} &=& \langle \psi_{0}^+ |\hat A^{m} \hat z\hat
A^{+n}|\psi_{0}^{-}\rangle
 =   \langle \psi_{0}^+ |\hat A^{m-1} [\hat A, \hat z]
         \hat A^{+n}|\psi_{0}^{-}\rangle
         + \langle \psi_{0}^+ |\hat A^{m-1}   \hat z\hat A
                      (\hat A^+)^{n}|\psi_{0}^{-}\rangle
                      \end{eqnarray}
and proceed as before. We find that $J_{mn}= 0$ for $m>n+1$. For
$m=n+1$ we obtain
\begin{eqnarray}
J_{n+1n} \!&=&\! \langle \psi_{0}^+ |\hat A^{n} |\hat z|^{-1}\hat a^+
         (\hat A^+)^{n}|\psi_{0}^{-}\rangle
         \!+\! \langle \psi_{0}^+ |\hat A^{n} \hat a^+|\hat z|^{-1}
         (\hat A^+)^{n}|\psi_{0}^{-}\rangle
             \!+\! \langle \psi_{0}^+ |\hat A^{n}
             |\hat z|\hat A
                      (\hat A^+)^{n}|\psi_{0}^{-}\rangle\nonumber\\
&=& \tilde J_{n+1n}^{(1)}+ \tilde J_{n+1n}^{(2)}+ \tilde J_{n+1n}^{(3)}
                      \end{eqnarray}
Consider first $\tilde J_{n+1n}^{(3)}$:
\begin{equation}
\tilde J_{n+1n}^{(3)} = (C_{n}^-)^2\tilde J_{nn-1}^{(3)}
                      = 3n (3n+2)\tilde J_{nn-1}^{(3)}\,.
\end{equation}
Second, using
$\hat A^{n} |\hat z|^{-1}\hat a^+ (\hat A^+)^{n}
=\hat A^{n-1} \hat a^+|\hat z|^{-1} \hat A\hat (\hat A^+)^{n}$
we determine $\tilde J_{n+1n}^{(1)}$:
\begin{equation}
\tilde J_{n+1n}^{(1)} = (C_n^-)^2\tilde J_{nn-1}^{(2)}
                      =3n(3n+2)  \tilde J_{nn-1}^{(2)}\,.
\end{equation}
Third,
\begin{eqnarray}
\tilde J_{n+1n}^{(2)} &=& (\lambda_n^+-1)(\lambda_{n-1}^--1)\tilde
J_{nn-1}^{(2)}
= (3n+1)3n\,J_{nn-1}^{(2)}\,.
\label{eq:71}
\end{eqnarray}
We deduce that
\begin{equation}
\tilde J_{n+1n}^{(1)} =\frac{3n+2}{3n+1} \tilde J_{n+1n}^{(2)}
\end{equation}
and obtain the recursion
\begin{eqnarray}
J_{n+1n} &=& \tilde J_{n+1n}^{(1)}
            +\tilde J_{n+1n}^{(2)} +\tilde J_{n+1n}^{(3)} \label{eq:r4}
       = 3n(3n+2) J_{nn-1} + \bigg(1+\frac{3n+2}{3n+1}\bigg)
       \tilde J_{n+1n}^{(2)} \,.
       \end{eqnarray}
Iterating (\ref{eq:71}) we obtain

\begin{eqnarray}
\tilde J_{n+1n}^{(2)} &=& \left(\prod_{k=1}^n 3k(3k+1)\right) \tilde J_{10}^{(2)}
= \frac{3\sqrt{3}}{2\pi} 9^n\Gamma(2/3)\Gamma(n+1)\Gamma(n+4/3) J_{00}
\end{eqnarray}
and thus from  (\ref{eq:r4})
\begin{equation}
\label{eq:75}
J_{n+1n} = \frac{\sqrt{3}}{2\pi} 9^n \Gamma(2/3)\Gamma(2+n)\Gamma(n+4/3)\,
J_{00}\,.
\end{equation}
Comparing this result to (\ref{eq:Jmn}) we find
that (\ref{eq:Jmn}) gives the correct result for $m=n+1$,
although it was derived assuming $m\leq n$.
Normalising to obtain $Z_{mn}$ our final result is
\begin{equation}
\label{eq:87}
Z_{mn} = (-1)^{m-n} \frac{3^{5/6}}{6\pi}(m+n+1) \Gamma(2/3)
\frac{\sqrt{\Gamma(n+1)\Gamma(m+1/3)}}{\sqrt{\Gamma(m+1)\Gamma(n+5/3)}}
\frac{\Gamma(n-m+1/3)}{\Gamma(n-m+2)}
\end{equation}
for $n\geq m-1$ and zero otherwise.

\subsection{Ratios of eigenfunctions}
\label{sec:w}
In this section we show how to evaluate $\psi_n^+(0)/\psi_0^+(0)$.
For $z>0$, the eigenfunctions
 $\psi^+_n(z)$ are of the form $N_n^+ g_n(z)\exp(-z^3/6)$,
 where $g_n(z)$ is polynomial in $z^3$, of the form
 \begin{equation}
 g_n(z)=g_n^{(0)}+g_n^{(1)}z^3+\ldots
 \end{equation}
We determine how $\hat A^+$ and $\hat H$ act on these polynomials.
To this end we define
\begin{eqnarray}
{\protect\hat A}^{'+} &=& {\rm e}^{z^3/6}\hat A^+ {\rm e}^{-z^3/6} =
(\partial_z-z^2) z^{-1} (\partial_z-z^2)\\
{{\hat H}'} &=& {\rm e}^{z^3/6}\hat H {\rm e}^{-z^3/6} =
(\partial_z-z^2) z^{-1} \partial_z\,.
\end{eqnarray}
This implies
\begin{equation}
{\protect\hat A}^{'+}-{{\hat H}'}=-(\partial_z -z^2)z
=z^3-z\partial_z-1
\end{equation}
and thus
\begin{equation}
\label{eq:Ap}
(\hat A^{'+}-\hat H')g_n=-g_n^{(0)}+O(z^3)\,.
\end{equation}
Using $\lambda_n^+ = -3n$ we obtain
\begin{equation}
\label{eq:ev}
H'g_n=-3ng_n^{(0)}+O(z^3)
\end{equation}
from the eigenvalue equation.
Taking (\ref{eq:Ap}) and (\ref{eq:ev})  together we have
\begin{equation}
\label{eq:87a}
\hat A^{'+} g_n = -(3n+1) g_n^{(0)}+O(z^3)\,.
\end{equation}
Eq. (\ref{eq:87a})  implies
\begin{equation}
g_{n+1}^{(0)} = -(3n+1) g_n^{(0)}\,.
\end{equation}
With (\ref{eq:C}) it follows
\begin{eqnarray}
\psi_{n+1}^{+} (0) &=& N_{n+1}^+ g_{n+1}^{(0)}
  = -(3n+1) N_{n+1}^+/N_n^+  \psi_n^+(0)
= - \sqrt{\frac{3n+1 }{3(n+1)}}  \psi_n^+(0)\,.
  \label{eq:72}
\end{eqnarray}
Our final result for the ratio of wave-function amplitudes
is therefore
\begin{equation}
\label{eq:97}
\psi_n^+(0)/\psi_{0}^+(0) = (-1)^n\sqrt{\frac{\sqrt{3}\Gamma(2/3)}{2\pi}
\frac{\Gamma(n+1/3)}{\Gamma(n+1)}}\,.
\end{equation}

\section{Equilibrium correlations, diffusion and anomalous diffusion}
\label{sec:diff}
In this section  the momentum correlation function
in equilibrium and the time dependence of
$\langle x^2(t)\rangle$  are determined.

\subsection{Momentum correlation function in equilibrium}
The correlation function of momentum in equilibrium is
obtained from (\ref{eq:OO}). We have
\begin{equation}
\langle p_t p_{0}\rangle_{\rm eq.}
= p_0^2\left(\frac{D_1}{\gamma p_0^2}\right)^{2/3}
\sum_{n} Z_{0n}^2 \exp(\lambda_n^- t')\,.
\end{equation}
Using (\ref{eq:87}) we obtain
\begin{equation}
\label{eq:pcorrt}
\langle p_t p_{0}\rangle_{\rm eq.}
= \frac{\Gamma(4/3)}{3^{1/3}\Gamma(5/3)}
\left(\frac{p_0D_1}{\gamma}\right)^{2/3}
{\rm e}^{-2\gamma t} F_{21}\Big(\frac{1}{3},\frac{1}{3};\frac{5}{3};
{\rm e}^{-3\gamma t}\Big)
\end{equation}
for $t>0$.  Here $F_{21}$ is a hypergeometric function \cite{Abr}.
It follows that $\langle p_t p_0\rangle_{\rm eq.}$
decays as $\exp(-2 \gamma t)$ at large times as opposed
to $\exp(-\gamma t)$ in the Ornstein-Uhlenbeck process.

\subsection{Diffusion at long times}
We now turn to $\langle x^2(t)\rangle$. This expectation value
is calculated
using eqs. (\ref{eq:60}), (\ref{eq:61}), (\ref{eq:87}), and (\ref{eq:97}).
We have
\begin{eqnarray}
\label{eq:82}
\langle x^2(t)\rangle\!\!
&=& \!\!
\Big(\frac{p_0 D_1}{\gamma}\Big)^{2/3}\frac{1}{m^2 \gamma^2}
\sum_{k=0}^\infty\sum_{l=k-1}^\infty \frac{\psi_k^+(0)}{ \psi_0^+(0)} Z_{0l} Z_{kl}T_{kl}
\end{eqnarray}
with
\begin{eqnarray}
T_{kl}(t')&=&  \int_0^{t'} \!{\rm d}t_1'
         \int_{t_1'}^{t'} \!{\rm d}t_2' \,
{\rm e}^{\lambda_l^-(t_2'-t_1')+\lambda_k^+t_1'}
+\int_0^{t'} \!{\rm d}t_1'
         \int_{0}^{t_1'} \!{\rm d}t_2'\,
         {\rm e}^{\lambda_l^-(t_1'-t_2')+\lambda_k^+t_2'}
         \nonumber\\
 &=&2\frac{\lambda_k^+(1-{\rm e}^{\lambda_l^- t'})-\lambda_l^-(1-{\rm
 e}^{\lambda_k^+
 t'})}{\lambda_l^-\lambda_k^+(\lambda_k^+-\lambda_l^-)}\,.
\end{eqnarray}
We define
\begin{equation}
\label{eq:coeff}
A_{kl} \equiv \frac{\psi_k^+(0)}{ \psi_0^+(0)} Z_{0l} Z_{kl}
=\frac{3^{2/3}\Gamma(2/3)^2}{12\pi^2}
\frac{(k\!+\!l\!+\!1)\Gamma(k\!+\!1/3)\Gamma(l\!+\!1/3)\Gamma(l\!-\!k\!+\!1/3)}{\Gamma(k+1)\Gamma(l+5/3)\Gamma(l-k+2)}
\end{equation}
for $k \geq l-1$ and zero otherwise.
In order to determine $\langle x^2(t)\rangle$, the sum
\begin{equation}
\label{eq:82a}
S(t') = \sum_{kl} A_{kl} T_{kl}(t')
\end{equation}
is required.  Note that $A_{kl}=0$ for $k< l-1$.
Consider the behaviour of (\ref{eq:82a})  at large values
of $t'$. We write the $k=0$ term separately
\begin{equation}
T_{0l} = -\frac{2t'}{\lambda_l^-}+\frac{2}{{\lambda_l^-}^2} ({\rm e}^{\lambda_l^- t'}-1)\,.
\end{equation}
This gives
\begin{eqnarray}
\label{eq:85}
\langle x^2(t)\rangle &=& 2 {\cal D}_x t\\
\nonumber
&+&  \Big(\frac{p_0 D_1}{\gamma}\Big)^{2/3}\frac{1}{m^2 \gamma^2}
\sum_{l=0}^\infty  Z_{0l}^2 \frac{2}{{\lambda_l^-}^2} ({\rm e}^{\lambda_l^- t'}-1)\\
&+&  \Big(\frac{p_0 D_1}{\gamma}\Big)^{2/3}\frac{1}{m^2 \gamma^2}
\sum_{k=1}^\infty\sum_{l=k-1}^\infty \frac{\psi_k^+(0)}{ \psi_0^+(0)} Z_{0l} Z_{kl}T_{kl}(t')\nonumber\,.
\end{eqnarray}
At large $t'$ the secular term dominates
and diffusion is thus recovered.
The diffusion constant is obtained as \cite{Arv06}
\begin{equation}
\label{eq: 55}
{\cal D}_x\!=\frac{(p_0D_1)^{2/3}}{m^2\gamma^{5/3}}
\frac{\pi 3^{-5/6}}{2\,\Gamma\!(2/3)^2} F_{32}
\Big(\frac{1}{3},\frac{1}{3},\frac{2}{3};\frac{5}{3},\frac{5}{3};1\Big)\,.
\end{equation}

\subsection{Anomalous diffusion at short times}
Next we consider short times. In order to evaluate (\ref{eq:82a})
at small values of $t'$, we replace the sums in (\ref{eq:82a}) by
integrals:
\begin{equation}
\label{eq:int}
S(t') =  \int_0^\infty \!\!{\rm d}l \int_0^l \!\!{\rm d}k\ T(k,l,t)A(k,l)\,.
\end{equation}
In order to evaluate (\ref{eq:int})
we use the asymptotic form of the coefficients $A_{kl}$:
\begin{equation}
A(k,l) \sim  \frac{3^{2/3}\Gamma(2/3)^2}{12\pi^2}\frac{k+l}{k^{2/3}
l^{4/3} (l-k)^{5/3}}\,.
\end{equation}
The coefficients $A(k,l)$ exhibit
non-integrable divergence $k\to l$. In view of this divergence
we make use of a sum rule of  the $A_{kl}$:
\begin{equation}
\label{eq:86}
\sum_{k=0}^{l+1} A_{kl}=0 \,.
\end{equation}
It can be derived by considering
\begin{equation}
\sum_{k=0}^{l+1}  \psi_k^+(z) \langle\psi_k^+|\hat z|\psi_l^-\rangle
= \sum_{k=0}^\infty  \psi_k^+(z) \langle\psi_k^+|\hat z|\psi_l^-\rangle
= \sum_{k=0}^\infty  \langle z|\psi_k^+\rangle \langle\psi_k^+|\hat
z|\psi_l^-\rangle
=\langle z |\hat z|\psi_l^-\rangle
\end{equation}
which vanishes for $z=0$. Replacing sums by integrals the sum rule
amounts to
\begin{equation}
\label{eq:103}
\int_0^l\! {\rm d}k\ A(k,l)=0\,.
\end{equation}
Eq. (\ref{eq:103})  allows us to write
\begin{equation}
\label{eq:104}
S(t')=\int_0^\infty\!\! {\rm d}l \int_0^l\!\! {\rm d}k\
\biggl[T(k,l;t')-\lim_{k\to l}T(k,l;t')\biggr]
A(k,l)\,.
\end{equation}
We find that the divergence of $A(k,l)$ is reduced to an
integrable divergence by the fact that $T(k,l;t)-\lim_{k\to
l}T(k,l;t)=O(k-l)$. Approximately, $T(k,l;t)$ is given by
\begin{equation}
T(k,l;t)=-2\,{3k[1-\exp(-3lt)]-3l[1-\exp(-3kt)]\over{27lk(l-k)}}\,.
\end{equation}
Changing the integration variables in (\ref{eq:104}) to $x=3lt$
and $xy=3kt$,  we have $k=xy/(3t)$ and $l=x/(3t)$, and
\begin{equation}
A(x,y,t)=\frac{3^{10/3}\Gamma(2/3)^2}{12\pi^2}t^{8/3}x^{-8/3}
\frac{1+y}{y^{2/3}(1-y)^{5/3}}.
\end{equation}
The Jacobian of the transformation is $J=x/(9t^2)$.
In the new variables,
\begin{equation}
T(x,y;t)=-2\,{t^2\over{x}}\biggl[{a(x)-a(xy)\over{1-y}}\biggr]
\end{equation}
where $a(x)=[1-\exp(-x)]/x$. Furthermore
\begin{equation}
\lim_{y\rightarrow
1}T(x,y;t)=-2\frac{t^2}{x}\lim_{y\rightarrow
1}\left[\frac{a(x)-a(xy)}{1-y}\right]\,.
\end{equation}
Using
\begin{equation}
\lim_{y\rightarrow 1}T(x,y;t)=-2\frac{t^2}{x}\left[
\frac{\partial}{\partial y}a(xy)\right] \Bigg|
_{y=1}=-2\frac{t^2}{x} x a'(x)
\end{equation}
we finally obtain
\begin{eqnarray}
T(k,l;t)-\lim_{k\to l}T(k,l;t)&=&-2\,{t^2\over{x}}
\biggl[{a(x)-a(xy)\over{1-y}}-xa'(x)\biggr].
\end{eqnarray}
This gives
\begin{eqnarray}
\label{eq:an}
S(t)&=& -C\,{t}^{8/3}\int_0^\infty\!\! {\rm d}x\ x^{-8/3} \int_0^1
\!\!{\rm d}y\ \biggl[{a(x)-a(xy)\over{1-y}}-xa'(x)\biggr]
{1+y\over{y^{2/3}(1-y)^{5/3}}}
\end{eqnarray}
with $C = 3^{1/3}\Gamma(2/3)^2/(2\pi^2)$.
Equation (\ref{eq:an}) implies anomalous spatial diffusion at
short times:
\begin{equation}
\label{eq:x2}
\langle x^2(t)\rangle = {\cal C}_x \,\bigl((p_0
D_1)^{2/3}m^{-2}\bigr)\,{t}^{8/3}
\end{equation}
with
\begin{equation}
\label{eq:cx1}
{\cal C}_x = -C\,\int_0^\infty {\rm d}x\ x^{-8/3} \int_0^1 {\rm d}
y\ \biggl[{a(x)-a(xy)\over{1-y}}-xa'(x)\biggr]
{1+y\over{y^{2/3}(1-y)^{5/3}}}\,.
\end{equation}

This anomalous diffusion is analogous to that described by
Golubovic, Feng and Zeng \cite{Gol91} in the case where there is
no damping (they considered the case where $\zeta=3$ in their
paper). Our results give the prefactor as well as the scaling
behaviour. It is noteworthy that we are able to solve the problem
discussed in \cite{Gol91} by solving a more complex set of
equations exactly, and taking a limit.

\section{WKB analysis}
\label{sec:wkb}

The staggered ladder spectrum discussed in section
\ref{sec:specdec} is surprising, especially in view of the fact
that for large quantum number $n$ we expect that the
eigenfunctions of the Hamiltonian (\ref{eq:H}) might be obtained
by WKB theory. 
In this section we show how to obtain the spectrum (\ref{eq:32})
of $\hat H$ using asymptotic Wentzel-Kramers-Brillouin (WKB)
analysis.  We show how phase shifts associated with the
singularity at $z=0$ of the Hamiltonian (\ref{eq:H}) are the
source of the staggered spectrum. It turns out that the WKB
procedure gives rise to the exact eigenvalues.

\begin{figure}
\includegraphics[width=5cm,clip]{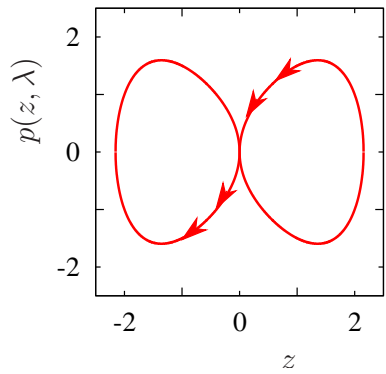}
\caption{\label{fig:2} The trajectories of the classical
Hamiltonian (\ref{eq:Hcl}) are figure-of-eight orbits.}
\end{figure}

In dimensionless coordinates, the classical Hamilton function
corresponding to (\ref{eq:H}) is
\begin{equation}
\label{eq:Hcl}
H_{\rm cl} = \frac{1}{2} - \frac{z^3}{4} -  p^2/z\,.
\end{equation}
Solving $H_{\rm cl}=\lambda$ for $p$ we obtain
$p(z,\lambda) = \pm \frac{1}{2}\sqrt{(2-4 \lambda)z-z^4}$, while the
velocity is
\begin{equation}
\dot z = \partial H_{\rm cl}/\partial p = -2 p/z\,.
\end{equation}
The classical trajectories are figure-of-eight orbits,
illustrated in figure \ref{fig:2}.

The WKB wavefunction is of the form
\begin{equation}
f(z) = \big[z/p(z,\lambda) \big]^{1/2} \exp\left(\pm {\rm i} \int^{z}\! {\rm d}z\, p(z,\lambda)\right).
\end{equation}
The phase of the wave function can be determined as follows.
We discuss separately the behaviours of the wavefunction
at the origin $z_0=0$ and in the vicinity of the regular turning
point $z_{\rm t.p.}= (2-4\lambda)^{1/3}$. In the latter case,
the wavefunction at $z < z_{\rm t.p.}$
connected with the turning point is
\begin{equation}
\label{eq:z1}
f_{\rm t.p.}(z) = C_{\rm t.p.}\big[z/p(z,\lambda)\big]^{1/2}
\sin\left(\int_z^{z_{\rm t.p.}}\! {\rm d}z\, p(z,\lambda)+\frac{\pi}{4}\right)\,.
\end{equation}

Consider now the behaviour of the wavefunction near the origin.
The Hamiltonian has a singularity at $z=0$. We find an exact
solution of the equation when $\gamma=0$. This equation has a
continuous spectrum, but we identify solutions $f_+(z)$ and
$f_-(z)$ which correspond respectively to even and odd solutions
of the full equation for $\gamma >0$ (with discrete spectrum).
Close to $z=0$, the damping term in the Hamiltonian is negligible,
and for $z>0$ the eigenfunctions resemble solutions of the
equation
\begin{equation}
\partial_z z^{-1}\partial_z f(z)=-\Lambda f(z)
\end{equation}
where $\Lambda(=-\lambda)$ is a positive constant. Write $f=F'$,
and find that
\begin{equation}
\frac{\partial}{\partial z} \biggl({F''+\Lambda zF\over{z}}\biggr)=0
\end{equation}
so that $F''+\Lambda zF=Cz$ for some constant $C$. Thus we find
$G(z)=F(z)-C/\Lambda$ satisfies $G''+z\Lambda G=0$, which has
solution $G(z)={\rm Ai}(-\Lambda^{1/3}z)$, and a similar solution
constructed from ${\rm Bi}(x)$ (here Ai$(y)$ and Bi$(y)$ are Airy
Ai and Bi functions \cite{Abr}). The general solution is
\begin{equation}
\label{eq:Ai}
f(z)=A_1{\rm Ai}'(-\Lambda^{1/3}z)+A_2{\rm Bi}'(-\Lambda^{1/3}z) .
\end{equation}
We must find solutions of this form which resemble
the behaviour of the eigenfunctions of the equation
with $\gamma >0$ which obey the boundary conditions
\begin{equation}
\frac{{\rm d}^2f_+(0)}{{\rm d}z^2}= 0 \quad
\mbox{and}\quad f_-(0) = 0\,.
\end{equation}
The functions $\mbox{Ai}'$ and $\mbox{Bi}'$
have the following forms in the neighbourhood
of $z=0$
\begin{eqnarray}
\mbox{Ai}'(y) &=& c_1 \Big(\frac{y^2}{2}+{\cal O}(y^5)\Big)
-c_2 \Big(1+\frac{y^3}{3}+{\cal O}(y^6)\Big)\\
\mbox{Bi}'(y) &=& \sqrt{3} c_1 \Big(\frac{y^2}{2}+{\cal O}(y^5)\Big)
+\sqrt{3} c_2 \Big(1+\frac{y^3}{3}+{\cal O}(y^6)\Big)
\end{eqnarray}
with $y = -\Lambda^{1/3} z$, $c_1 = 3^{-2/3}/\Gamma(2/3)$,
and  $c_2 = 3^{-1/3}/\Gamma(1/3)$. So the positive-parity
solution corresponds to the choice
\begin{equation}
A_1^+ = -\sqrt{3}\quad\mbox{and}\quad A_2^+ = 1
\end{equation}
while
\begin{equation}
A_1^- = \sqrt{3}\quad\mbox{and}\quad A_2^- = 1\,.
\end{equation}
At large values of $y$ the corresponding wave functions are of the form
\begin{equation}
f_\pm(y) \sim (-y)^{1/4} \sin\Big(\frac{2}{3} (-y)^{3/2} +\frac{\pi}{4}
\pm \frac{\pi}{3}\Big)\,.
\end{equation}
Noting that, near $z=0$,  $\int_0^{z}\! {\rm d}z\, p(z,\lambda) =
(2/3) (-y)^{3/2}$ and $(z/p(z,\lambda))^{1/2} \propto
(-y)^{1/4}$, the solution coming from the origin is
\begin{equation}
\label{eq:z0}
f_{\rm O,\pm}(z)=C_{\rm O,\pm}\big[z/p(z,\lambda)\big]^{1/2}
\sin\Big(\int_0^{z}\! {\rm d}z\, p(z,\lambda)+\frac{\pi}{4}\pm
\frac{\pi}{3}\Big)\,.
\end{equation}
The forms (\ref{eq:z1}) and (\ref{eq:z0}) should be smoothly
connected for $0 < z < z_{\rm t.p.}$. This requires
\begin{equation}
\int_0^{z_{\rm t.p.}}\! {\rm d}z\, p(z,\lambda)+ \frac{\pi}{2} \pm \frac{\pi}{3}  = (n+1)\pi,
\end{equation}
for $n = 0,1,\ldots$ together with $C_{\rm t.p.}=(-1)^n C_{\rm O,\pm}$.
Writing
\begin{equation}
\label{eq:action} S(\lambda)=\int_0^{z_{\rm t.p.}}\! {\rm d}z\,
p(z,\lambda)= \frac{\pi}{12}(-4\lambda+2)\,,
\end{equation}
we find the quantisation condition
\begin{equation}
\label{eq:qtcd}
S(\lambda_\pm) = \Big(n+\frac{1}{2}\mp \frac{1}{3}\Big)\pi\,.
\end{equation}
Using (\ref{eq:action}) the quantisation condition takes the form
\begin{equation}
\lambda = \left\{\begin{array}{ll}
-3n & \mbox{even parity}\\
-3n-2 & \mbox{odd parity}
\end{array}\right .
\end{equation}
which corresponds exactly to the spectrum (\ref{eq:32}) obtained
by exactly diagonalising $\hat H$.

\section{Results for other values of $\zeta$}
\label{sec:gen}

Up to now we have only considered the case of generic random
forcing (where the constant $D_1$ in equation (\ref{eq: 3}) is not 
zero), corresponding to the case $\zeta=1$ in equation (\ref{eq:
4}). In this section we explain two cases where other values of
$\zeta$ arise and briefly describe results for arbitrary
positive values of $\zeta$, analogous to the results obtained in section 6.

First consider the case where the force is the gradient of a
potential, $f(x,t)=\partial V(x,t)/\partial x$. We assume that the
potential has mean value zero and correlation function ${\cal
C}(X,T)=\langle V(x+X,t+T)V(x,t)\rangle$. Assuming that ${\cal
C}(X,T)$ is sufficiently differentiable at $T=0$, the diffusion
constant is
\begin{eqnarray}
\label{eq:expdp}  D(p)&=&{1\over{2}}\int_{-\infty}^\infty {\rm
d}t\ \biggl\langle {\partial V\over{\partial x}}(pt/m,t){\partial
V\over{\partial x}}(0,0)\biggr\rangle
\nonumber\\
&=&{-m\over{2\vert p\vert}}\int_{-\infty}^\infty {\rm d}X\
{\partial^2 {\cal
C}\over{\partial X^2}}(X,mX/p)\nonumber \\
&=&{-m\over{2\vert p\vert}}\int_{-\infty}^\infty {\rm d}X\
\biggl[{\partial^2{\cal C}\over{\partial x^2}}(X,0)+{m\over
p}X{\partial^3 {\cal C}\over {\partial^2 X\partial
T}}(X,0)\nonumber \\
&&\ \ \ +{m^2\over{2p^2}}X^2{\partial^4 {\cal C}\over{\partial^2
X\partial^2 T}}(X,0)+O(X^3)\biggr] \ .
\end{eqnarray}
Integration by parts shows that the integral over the first term
of the expansion is zero, and the integral over the second term is
zero by symmetry. The leading-order contribution in $\vert
p\vert^{-1}$ comes from the third term. Integrating this term by
parts twice gives
\begin{equation}
D(p)\sim {-m^3\over{2\vert p\vert^3}}\int_{-\infty}^\infty {\rm
d}X\ {\partial^2 {\cal C}\over{\partial T^2}}(X,0)\ .
\end{equation}
Thus in the case of a potential force with a sufficiently smooth
correlation function we have $\zeta=3$.

An exceptional case which is worthy of comment is when the
potential $V(x,t)$ is itself generated from a set of
Ornstein-Uhlenbeck processes $A_j(t)$ by writing $V(x,t)=\sum_j
A_j(t)\Phi_j(x)$, where the $\Phi_j(x)$ are elements of some
suitable set of basis functions. In this case the correlation
function of $V(x,t)$ is of the form $c(x)\exp(-\gamma \vert
t\vert)$ [for some function $c(x)$]. Then the second term in the
expansion on the final line of equation (\ref{eq:expdp}) does not
vanish by symmetry and we find $D(p)\propto \vert p\vert^{-2}$,
that is $\zeta=2$.

For general positive values of $\zeta$ the Hamiltonian
(\ref{eq:3}) is replaced by
\begin{equation}
\hat H = \frac{1}{2} - \frac{1}{4}|z|^{2+\zeta}
+{\partial\over{\partial z}}\frac{1}{|z|^\zeta}{\partial\over{\partial z}}\,.
\end{equation}
Its ground state
\begin{equation}
\lambda_0^+ = 0\,\quad\mbox{and}\quad
\psi_0^+(z) = {\cal C}_0^+ {\rm e}^{-|z|^{\zeta+2}/(4+2\zeta)}
\end{equation}
and first excited state
\begin{equation}
\lambda_0^- = -1-\zeta\,\quad\mbox{and}\quad
\psi_0^-(z) = {\cal C}_0^- z |z|^\zeta\,{\rm e}^{-|z|^{\zeta+2}/(4+2\zeta)}
\end{equation}
are found by inspection.
Raising and lowering operators can be introduced in a manner
analogous to eqs. (\ref{eq:ham}-\ref{eq:gam}). We write
\begin{equation}
\hat H = \hat a^- |z|^{-\zeta} \hat a^+
\end{equation}
with $\hat a^\pm = \partial_z\pm z|z|^\zeta/2$. The operators
\begin{equation}
\hat A = \hat a^+ |z|^{-\zeta} \hat a^+\,\quad\mbox{and}\quad
\hat A^+ = \hat a^- |z|^{-\zeta} \hat a^-
\end{equation}
satisfy
\begin{equation}
[\hat H, \hat A] = (2+\zeta)\hat A\quad \mbox{and}\quad
[\hat H, \hat A^+] = -(2+\zeta)\hat A^+\,.
\end{equation}
and act as lowering and raising operators.
For the spectrum of $\hat H $ we obtain
\begin{equation}
\lambda_n^+ = -(2+\zeta)\,n\quad\mbox{and}\quad\lambda_n^- = -(2+\zeta)n-1-\zeta\,.
\end{equation}
These expressions replace (\ref{eq:32}).
Note also that the commutator of $\hat A$ and $\hat A^+$ is
\begin{equation}
[\hat A, \hat A^+] = -(2+\zeta) (\hat H + \hat G)
\end{equation}
where $\hat G = \hat a^+ |z|^{-\zeta} \hat a^-$ and
$\hat H-\hat G = \hat I$. The normalisation of the eigenstates
\begin{eqnarray}
\hat A^+ |\psi_{n}^-\rangle &=& C_{n+1}^-  |\psi_{n+1}^- \rangle \\
\hat A   |\psi_{n}^-\rangle&=& C_{n\phantom{+1}}^-  |\psi_{n-1}^-
\rangle
\end{eqnarray}
is determined as in section \ref{sec:eigenstates}. We obtain
\begin{equation}
(C_{n+1}^-)^2 =(2+\zeta)(n+1)[(2+\zeta) n+3+2 \zeta]
\end{equation}
and
\begin{equation}
(C_{n+1}^+)^2 =(2+\zeta)(n+1)[(2+\zeta)n+1]\,.
\end{equation}
The results of section \ref{sec:corr} for the matrix elements
$Z_{mn} = \langle \psi_m^-|\hat z|\psi_n\rangle$  and for
$\psi_n^+(0)/\psi_0^+(0)$ generalise
as follows:
\begin{equation}
Z_{mn}=(-1)^{n-m}
\frac{(2+\zeta)^{-\frac{1+\zeta}{2+\zeta}}}{\Gamma(\frac{\zeta}{2+\zeta})}(m+n+1)
\frac{\Gamma(\frac{\zeta}{2+\zeta}-m+n)
\sqrt{\Gamma(n+1)\Gamma(\frac{1}{2+\zeta}+m)}}{\Gamma(2-m+n)
\sqrt{\Gamma(\frac{3+2\zeta}{2+\zeta}+n)\Gamma(m+1)}}\,,
\end{equation}
\begin{eqnarray}
\psi_{n}^+(0)/\psi_{0}^+(0) &=& (-1)^n
\sqrt{\frac{\Gamma[((2+\zeta)n+1)/(2+\zeta)]}{\Gamma(n+1)\Gamma(1/(2+\zeta))}}
\,.
\end{eqnarray}
This allows us to obtain, for example,  the diffusion constant
\begin{equation}
\label{eq:Dx} {\cal D}_x =\frac{1}{m^2}\left (
\frac{p_0^{2\zeta}D_{\zeta}^2}{\gamma ^{4+\zeta}}\right
)^{\frac{1}{2+\zeta}}\frac{(2+\zeta)^{-\frac{4+3\zeta}{2+\zeta}}\pi
F_{32}(\frac{\zeta}{2+\zeta},\frac{\zeta}{2+\zeta},\frac{1+\zeta}{2+\zeta};
\frac{3+2\zeta}{2+\zeta},\frac{3+2\zeta}{2+\zeta};1)}
{\sin(\frac{\pi}{2+\zeta})\Gamma(\frac{3+2\zeta}{2+\zeta})^2}
\end{equation}
describing the dynamics at large times.  Upon substituting $\zeta=0$, eq. (\ref{eq:Dx}) reproduces
the standard Ornstein-Uhlenbeck result, and it gives (\ref{eq: 55})
for $\zeta=1$.

For the short-time anomalous diffusion we obtain
\begin{equation}
\label{eq:144}
\langle x^2(t) \rangle = {\cal C}_x
((p_0^{\zeta}D_{\zeta})^{\frac{2}{2+\zeta}}m^{-2})t^{\frac{6+2\zeta}{2+\zeta}}
\end{equation}
with
\begin{equation}
\label{eq:Cx}
{\cal C}_x = -C \int _0 ^{\infty} {\rm d}x\;
x^{-\frac{6+2\zeta}{2+\zeta}} \int _0 ^1 {\rm d}y \left
[\frac{a(x)-a(xy)}{1-y}-xa'(x) \right
]\frac{1+y}{y^{\frac{1+\zeta}{2+\zeta}}(1-y)^{\frac{4+\zeta}{2+\zeta}}},
\end{equation}
where $a(x)$ is the same as in (\ref{eq:cx1}), and
\begin{equation}
C=\frac{2(2+\zeta)^{-\frac{2\zeta}{2+\zeta}}}{\Gamma(\frac{\zeta}{2+\zeta})^2}.
\end{equation}
This reproduces (\ref{eq:x2}) for $\zeta=1$ and concludes
our summary of results for other than generic random
forcing.

{\em Acknowledgements.} BM acknowledges financial support from
Vetenskapsr\aa{}det. BM, MW, and KN thank Marko Robnik for
inviting them to the 2005 summer school in Maribor where this work
was initiated.

\newpage

\end{document}